\documentclass[twocolumn,superscriptaddress,nobibnotes,nofootinbib,floatfix,aps,prl,citeautoscript,reprint]{revtex4-1}

\usepackage{siunitx}
\usepackage{graphicx}   

\usepackage{graphicx}
\usepackage{epsfig}
\usepackage{subfigure}
\usepackage{color}
\usepackage{latexsym}
\usepackage{amsmath,bm}
\usepackage{dcolumn}           
\usepackage{natbib}
\usepackage[colorlinks,linkcolor=blue,citecolor=green]{hyperref}
\newcommand{\ub}{Department of Physics, Universit\"{a}t Basel,
Klingelbergstr. 82, 4056 Basel, Switzerland}
\newcommand{\nw}{Department of Materials Science and Engineering, Northwestern University, Evanston, IL 60208, USA}
\newcommand{\ucla}{Department of Materials Science and Engineering, University of California, Los Angeles, California 90095, USA}

\begin{document}

\title{Ultralow Thermal Conductivity in Full-Heusler Semiconductors}

\author{Jiangang He}
\thanks{These two authors contributed equally to this work}
\affiliation{\nw}

\author{Maximilian Amsler}
\thanks{These two authors contributed equally to this work}
\affiliation{\nw}

\author{Yi Xia}
\affiliation{\ucla}

\author{S. Shahab Naghavi}
\affiliation{\nw}

\author{Vinay I. Hegde}
\affiliation{\nw}

\author{Shiqiang Hao}
\affiliation{\nw}

\author{Stefan Goedecker}
\affiliation{\ub}

\author{Vidvuds Ozoli\c{n}\v{s}}
\affiliation{\ucla}

\author{Chris Wolverton}
\email{c-wolverton@northwestern.edu}
\affiliation{\nw}

\date{\today}

\begin{abstract}
Semiconducting half- and, to a lesser extent,
full-Heusler compounds are promising thermoelectric materials due to their compelling electronic properties with large power factors. However, intrinsically high thermal conductivity resulting in a limited thermoelectric efficiency has so far impeded their widespread use in practical applications. Here, we report the computational discovery of a class of hitherto unknown stable semiconducting full-Heusler compounds with ten valence electrons ($X_2YZ$, $X$=Ca, Sr, and Ba; $Y$= Au and Hg; $Z$=Sn, Pb, As, Sb, and Bi) through high-throughput {\sl ab-initio} screening. These new compounds exhibit ultralow lattice thermal conductivity $\kappa_{\text{L}}$ close to the theoretical minimum due to strong anharmonic rattling of the heavy noble metals, while preserving high power factors, thus resulting in excellent phonon-glass electron-crystal materials. 
\end{abstract}

\maketitle

Limited fossil resources and environmental challenges have sparked an intense search for alternate methods for clean, sustainable and efficient power generation in the last decade, with an increasing focus on thermoelectric (TE) materials for the direct conversion of electricity from (waste) heat~\cite{snyder_complex_2008}. The
conversion efficiency is measured by the dimensionless figure of merit $ZT=S^2 \sigma T/(\kappa_\text{e}+\kappa_\text{L})$, where $T$ is the absolute temperature, $S$ is the thermopower, $\sigma$ is the electrical conductivity, and $\kappa_\text{e}$ and $\kappa_{\text{L}}$ are the electronic and lattice thermal conductivities, respectively~\cite{rowe2005thermoelectrics}. Optimizing $ZT$ thus involves the tuning of conflicting materials properties by maximizing the power factor (PF) $\sigma S^2$ and minimizing thermal conductivity ($\kappa_{\text{e}}$ and $\kappa_{\text{L}}$) simultaneously. The former is commonly achieved by band tuning and heavy doping, while the latter involves structural engineering through disorder (such as alloying Bi$_2$Te$_3$~\cite{wright_thermoelectric_1958}), complexity (such as Zn$_4$Sb$_3$~\cite{snyder2004, bjerg_enhanced_2011, bjerg_modeling_2014,tadano_impact_2015}), nano-structuring (such as PbTe~\cite{wu_broad_2014}) or sub-structuring (skutterudites and clathrates~\cite{uher_chapter_2001,nolas_chapter_2001,nolas_recent_2006}), anharmonicity (Cu$_{12}$Sb$_4$S$_{13}$ \cite{Lu2013}), and ferroelectric-like lattice instability (SnSe \cite{SnSe,doi:10.1038/nphys3492}).

Semiconducting half-Heusler (HH) compounds with $XYZ$ composition and 18 valence electrons per formula unit (f.u.) have been recently actively studied as promising TE materials due to their excellent electronic and mechanical properties and thermal stability. $ZT$ values as high as 1 have been reported for the $n$-type semiconductors $X$NiSn and $X$CoSb, $X$=\{Ti, Hf, Zr\}, and values exceeding 1 for $p$-type Fe$Y$Sb, $Y$=\{V, Nb\} at temperatures of about 1000~K (see Ref.~\onlinecite{zhu_high_2015} and references therein). The source of their compelling thermoelectric efficiency lies in the extremely high PF due to both the intrinsically narrow band gaps, which gives rise to a large $\sigma$, and the sharp increase in the density of states around the Fermi level, enhancing $S$ according to Mott's theory~\cite{rao_properties_2006}. However, intrinsically high lattice thermal conductivity prevents a further improvement of $ZT$ in HH materials. Although alloying and nano-structuring are actively used to increase phonon scattering, it has so far not been possible to reduce the thermal conductivity to values below $\kappa_\text{L}= 6.7$~Wm$^{-1}$K$^{-1}$ at room temperature~\cite{Chen2013387}. In a recent high throughput \textit{ab-initio} study on 450 HH semiconductors merely four compounds were predicted to have $\kappa_{\text{L}}$ of less than 5~Wm$^{-1}$K$^{-1}$~\cite{PhysRevX.4.011019}.

In contrast to HH compounds, only a few full Heusler (FH) materials with $X_2YZ$ composition are semiconducting and have thus attracted limited attention as TE materials~\cite{Heusler,PhysRevB.66.174429}. Although very high PFs at room temperature have been reported in FH compounds based on Fe$_2$VAl with values as high as $4 - 6$~mW/m$^{-1}$K$^{-2}$~\cite{nishino_thermal_2006,vasundhara_electronic_2008,skoug_high_2009}, the effective $ZT$ values are merely around $0.13-0.2$ at 300~K~\cite{nishino_thermal_2006,terazawa_effects_2011,mikami_thermoelectric_2012} due to the high thermal conductivity (intrinsic $\kappa_\text{L}=28$~W/m$^{-1}$K$^{-1}$~\cite{nishino_thermal_2006}). Recently, Bilc~\textit{et al.} predicted several FH compounds through bi-elemental substitution in Fe$_2YZ$, thereby demonstrating the possibility of boosting the PF by a factor of 5 through band engineering~\cite{bilc_low-dimensional_2015}. However, similar to HH compounds, the high thermal conductivity has so far prevented the use of known FH materials in efficient TE applications.

In conventional FH compounds, late transition metals commonly form the $X_2$ simple cubic sublattice and early transition metals and \{III, IV, V\} main group elements are in the $Y$-$Z$ rock salt sublattice. It is well known that compounds with 24 valence electrons per f.u. can be semiconductors with fully occupied bonding states and empty anti-bonding states~\cite{Felser11}. The electronic structure of such materials can be readily explained by orbital hybridization and crystal field splitting \cite{PhysRevB.66.174429}. However, another FH system that can be semiconducting and contains 8 valence electrons per f.u. has been widely overlooked. These compounds do not contain transition metals and only have four fully occupied valence bands,
e.g. Li$_2$NaSb and K$_2$CsSb~\cite{felser_heusler_2015}.

To identify potential candidate materials with improved TE properties, we performed a high-throughput \textit{ab-initio} search for thermodynamically stable FH compounds with finite band gaps through elemental substitution within the qpmy~\cite{qmpy} framework in a chemical search space with 53 distinct elements (see~\cite{suppl}). We discovered a class of stable semiconducting FH compounds with intrinsically high PFs and extremely-low lattice thermal conductivity due to atomic rattling, which we call R-Heuslers. These compounds, which to our knowledge have not yet been reported in literature, contain 10 valence electrons per f.u. with alkaline earth elements \{Ba, Sr, Ca\} in the $X$ sublattice, whereas $Y$ are noble metals \{Au, Hg\} and $Z$ are main group elements \{Sn, Pb, As, Sb, Bi\}.

Density functional theory (DFT) calculations were performed within the 
projector augmented wave~(PAW) formalism~\cite{PAW-Blochl-1994} as implemented in the VASP~\cite{VASP-Kresse-1996,VASP-Kresse-1999} code together with the Perdew-Burke-Ernzerhof (PBE) approximation~\cite{Perdew-PBE-1996} to the exchange correlation potential~\cite{suppl}.
Initially, the thermodynamic stability of all FH compounds was evaluated based on the convex hull taking into account all competing phases included in the Open Quantum Material Database (OQMD)~\cite{OQMD,OQMD2}. All R-compounds have  negative formation energies (-0.6~$\sim$~-0.9 eV/atom) and were found to lie on the convex hull
(see Table~S1).
However, even if a compound lies on the  hull, there is still a possibility that a different crystal structure at given composition is lower in energy. We thus performed structural searches using the Minima Hopping Method (MHM) as implemented in the Minhocao package~\cite{mhm1,mhm2}. 
MHM simulations were launched starting from the FH structure with up to 4 f.u. per cell in order to explore neighboring low-lying minima on the potential energy surface, visiting dozens of distinct minima per compound. Additionally, 18 intermetallic prototype structures were screened. Both searches did not yield any structures with energies lower than the Heusler phase for Ba containing compounds, indicating that they are thermodynamically stable. However, a distorted structure with $P21/m$ symmetry was found through the MHM which is favored for a few Sr and Ca compounds 
(see Table~S2 in~\cite{suppl}),
rendering them metastable. Since the energy differences between the $P21/m$ and Heusler structures are small we expect that all R-Heuslers are experimentally accessible, such that we will hereon focus solely on the Heusler phases. Furthermore, phonon calculations were carried out with the frozen phonon method as implemented in the Phonopy package~\cite{phonopy} to confirm that all Heusler phases are dynamically stable with no imaginary phonons 
(see Fig.~S2).

\begin{figure}[h]
	\setlength{\unitlength}{1cm}
	\includegraphics[width=0.9\columnwidth,angle=0]{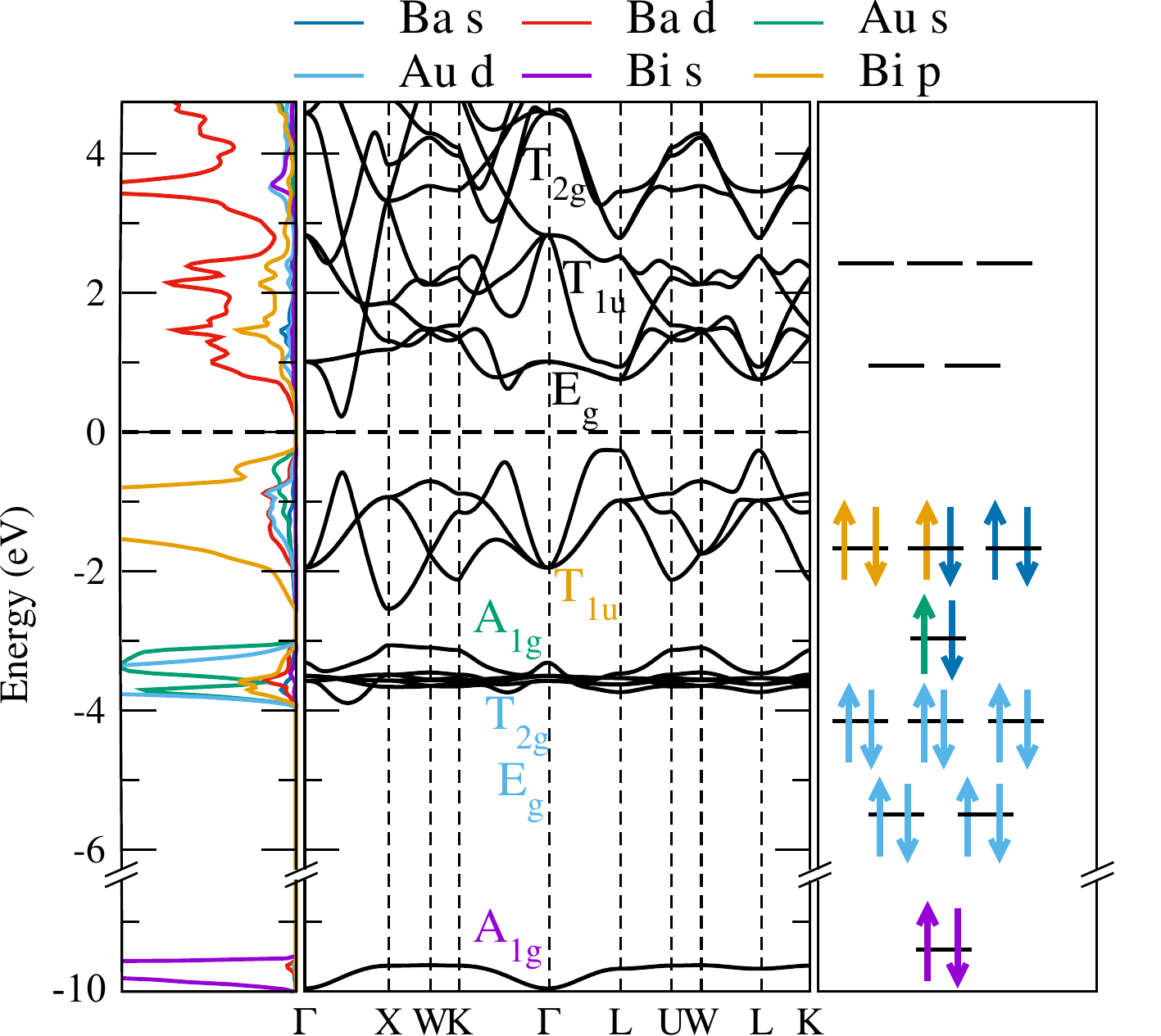}
	\caption{(color online) Band structure and density of states (DOS) of cubic full Heusler (Fm$\bar{3}m$) Ba$_2$AuBi computed at PBE level using WIEN2K code ~\cite{blaha_full-potential_1990} to extract the irreducible symmetry representation of each band at the $\Gamma$ point.}
	\label{bandstructure}
\end{figure}

Since semi-local DFT functionals are well known to underestimate  band gaps, the screened HSE06 hybrid functional~\cite{heyd_erratum:_2006,paier_erratum:_2006,heyd_hybrid_2003} was employed, leading to gaps between 0.01 (Ca$_2$AuBi) and 1.0~eV (Ba$_2$AuSb) 
(see Table~S1),
in the range of other good thermoelectrics. In Fig.~\ref{bandstructure} the band structure of Ba$_2$AuBi is shown as a representative case to illustrate how the R-Heuslers lead to finite band gap materials.
Near the Fermi level, the valence band is formed mainly by the Bi 6p orbitals and the conduction band is composed of the Ba 5d and Bi 6p* states. 
The two most electropositive atoms Ba donate their four 5s electrons to the  electronegative Au and Bi atoms. The fully occupied 5d and 6s states of the Au atom are extremely localized and far below the fermi level, such that Au only weakly interacts with its neighboring atoms. This weak bonding gives rise to its unique vibrational behavior as will be discussed later.
The Bi 6s lone pairs are buried deep below the Fermi level and are stereochemically inactive. Although the R-compounds have in total 10 valence electrons per f.u., these inactive Bi 6s lone pairs lead effectively to an 8-electron system, resulting in completely filled bonding and empty antibonding states in accordance with the electron counting rule to obtain a semiconductor.

\begin{figure}[b]
	\setlength{\unitlength}{1cm}
	\includegraphics[width=0.9\columnwidth,angle=0]{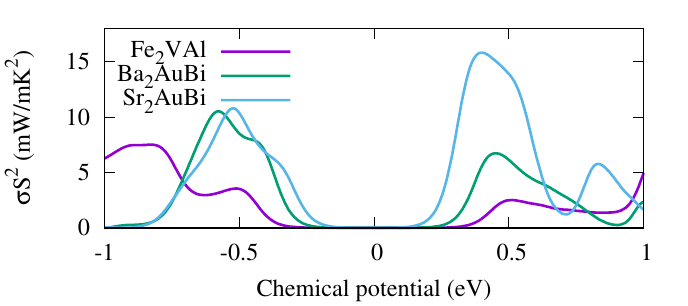}
	\caption{(color online) PFs of the three FH compounds Ba$_2$AuBi, Sr$_2$AuBi and Fe$_2$VAl at 300~K as a function of the electron chemical potential. Both n- and p-doping leads to significantly higher PFs of Ba$_2$AuBi and Sr$_2$AuBi compared to Fe$_2$VAl. For p-doping, Ba$_2$AuBi and Sr$_2$AuBi have improved PFs due to the ``flat-and-dispersive'' valence bands.}
	\label{fig:powerfactor}
\end{figure}

The electronic transport properties of the R-compounds were investigated to asses their quality as thermoelectric materials. Both the valence band maximum (VBM) and conduction band minimum (CBM) are at low symmetry points of the Brillouin zone. The VBM and CBM are located at the L and $\Delta$-point (between  $\Gamma$ and X) in the Brillouin zone, leading to high band degeneracy of 4 and 6, respectively. Furthermore, multiple bands have energies close to the VBM and CBM. Both properties lead to a sharp increase in the density of states around the Fermi level, which is favorable for large Seebeck coefficients. Bilc~\textit{et al.}~\cite{bilc_low-dimensional_2015} recently showed that bands which are flat along one direction and highly dispersive along others lead to high PFs in FH materials. In fact, such bands are present in the new R-compounds as well, where the valence band along L-$\Gamma$ is flat whereas it is highly dispersive along all other directions. To assess the thermoelectric conversion quality we solved the electronic Boltzmann transport equation within the constant relaxation time approximation using the Boltztrap code~\cite{Madsen200667}. The PBE band structures were resolved on a $41\times41\times 41$ k-point mesh, and a constant shift of the bands was applied to open the gaps to the values obtained with the more accurate HSE06 functional. Fig.~\ref{fig:powerfactor} compares the PFs of $X_2$AuBi ($X$=Ba and Sr) and the well studied thermoelectric compound Fe$_2$VAl. In accordance with the work of Bilc~\textit{et al.} a constant relaxation time of $\tau=3.4\times 10^{-14}$~s was used. For n-doped Fe$_2$VAl, the maximum value of $\sigma S^2$ lies at around 2.5~mWm$^{-1}$K$^{-2}$, which is in good agreement with experimental and theoretical values~\cite{bilc_low-dimensional_2015}. A value of 6~mWm$^{-1}$K$^{-2}$ is obtained for Ba$_2$AuBi at a similar carrier concentration. However, for p-type doping an even larger PF of 10~mWm$^{-1}$K$^{-1}$ in Ba$_2$AuBi can be observed due to the ``flat-and-dispersive'' band at the L-point.

Since all R-Heusler compounds are semiconductors, the dominating term in the thermal conductivity stems from the lattice vibrations. The accurate estimation of the lattice thermal conductivity requires the calculation of phonon scattering intensities. Only the anharmonic part of the lattice dynamics contributes to phonon-phonon interactions. The recently developed compressive sensing lattice dynamics~\cite{PhysRevLett.113.185501} (CSLD) technique was employed to obtain the third order force constants (FC). These FC were used to iteratively solve the linearized phonon Boltzmann equation with the ShengBTE package~\cite{ShengBTE_2014}. Cutoff radii of 12~\AA~and 6~\AA~were used to truncate the 2- and 3-body interactions, respectively, resulting in values of $\kappa_\text{L}$ converged to within 5\%.

\begin{figure}[h]
	\setlength{\unitlength}{1cm}
	\includegraphics[width=0.9\columnwidth,angle=0]{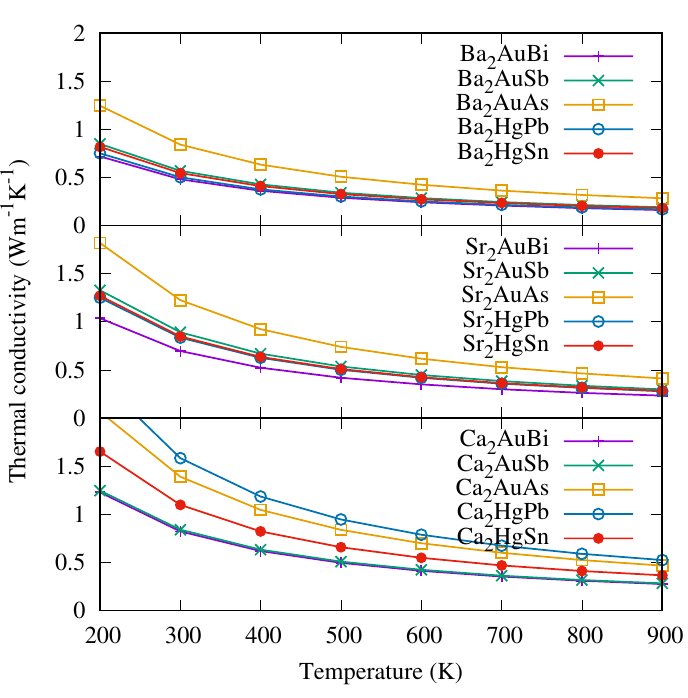}
	\caption{(color online) Lattice thermal conductivity of the various Heusler compounds as a function of temperature. Each panel contains the compositions Ba$_2$$YZ$, Sr$_2$$YZ$ and Ca$_2$$YZ$, respectively.}
	\label{fig:kappa}
\end{figure}

In Fig.~\ref{fig:kappa} the intrinsic lattice thermal conductivities for all R-compounds are shown as a function of temperature. All R-materials exhibit extremely low $\kappa_\text{L}$ (0.5~$\sim$~1.5~Wm$^{-1}$K$^{-1}$ at 300~K), roughly one order of magnitude lower than previous experimental~\cite{nishino_thermal_2006,Chen2013387} and theoretical~\cite{PhysRevX.4.011019} values in other FH and HH materials. The lowest $\kappa_\text{L}$ in Ba$_2$AuBi and Ba$_2$HgPb are even lower than 0.5~Wm$^{-1}$K$^{-1}$ at 300~K, which is comparable to the values along the out of plane direction of the van der Waals layers in SnSe ($\kappa_\text{L}=0.47$~Wm$^{-1}$K$^{-1}$), the material with the highest $ZT$  known to date~\cite{SnSe}. To verify the validity of the BTE approximation we computed the phonon mean free paths (MPF) at 300~K 
(see Fig.~S4),
indicating that the MPFs are overall longer than the smallest interatomic distances. However, we can expect that $\kappa_\text{L}$ will eventually converge to the amorphous limit of 0.27~W/m$^{-1}$K$^{-1}$ (Cahill-Pohl model \cite{10.1146/annurev.pc.39.100188.000521}) at higher temperatures. The discovery of such low values of $\kappa_\text{L}$ is all the more exciting since only complex crystal structures with large unit cells, e.g. skutterudites and clathrates, were believed to give rise to strong resonant scattering~\cite{toberer_phonon_2011}. Furthermore, we did not take into account additional phonon scattering mechanisms such as defects, dislocations, interfaces, and grain boundaries, which are known to additionally reduce the thermal conductivity. In fact, since FH and HH can be readily alloyed and nanostructured, we can expect a further reduction of $\kappa_\text{L}$ in structurally engineered samples of Ba$_2$AuBi and Ba$_2$HgPb.
\begin{figure}[h!]
	\setlength{\unitlength}{1cm}
	\includegraphics[width=0.9\columnwidth,angle=0]{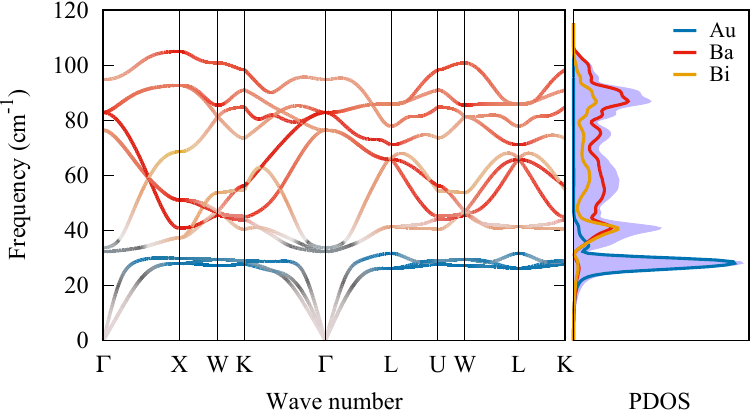}
	\caption{(color online) The phonon band structures are shown in the left panel. All bands are colored according to the amplitudes of the atomic eigendisplacements, where grey bands indicate equal contribution of all atoms. The PDOS is plotted in the right panel, with the shaded area indicating the DOS.}
	\label{fig:bandspdos}
\end{figure}

In order to understand the source of the extremely low lattice thermal conductivity we investigated the atomic contribution to the phonons and their density of states (DOS). Since all R-compounds have similar features in their phonon band structures 
(see Fig.~S2),
we will restrict our discussion to the material with the lowest lattice thermal conductivity, Ba$_2$AuBi.
The band structure shown in Fig.~\ref{fig:bandspdos} is colored according to the fractional amplitudes of the mode eigenvectors of each atom, indicating that flat, low-energy acoustic modes (around 25~cm$^{-1}$) are dominated by Au vibrations (blue), whereas optical branches are exclusively governed by Ba and Bi atoms (red and orange). In fact, there is barely any coupling between the acoustic branches and the optical modes with avoided-crossing, which is a clear indication of atomic rattling. Such strong rattling behavior with similar vibrational properties has so far only been observed in clathrates and skutterudites, which are well known for their low thermal conductivities where guest atoms rattle in the host cage structures and scatter the heat-carrying phonons~\cite{nolas_semiconducting_1998,cohn_glasslike_1999,blake_why_1999,dong_chemical_2000,dong_theoretical_2001,tse_phonon_2001,he_nanostructured_2014}. Analogously to clathrates, the low energy phonon branches can be interpreted as Au atoms rattling in a pseudo-cage structure composed of the alkali and main group atoms ($X$ and $Z$) 
(see Fig.~S3).
As discussed above, the strongly localized 6s state lead to weak interactions with the 6p states of the $Z$-site atoms, such that Au behaves like the inert alkaline (earth) metals in clathrate cages.

\begin{figure}[t]
	\setlength{\unitlength}{1cm}
	\includegraphics[width=0.8\columnwidth,angle=0]{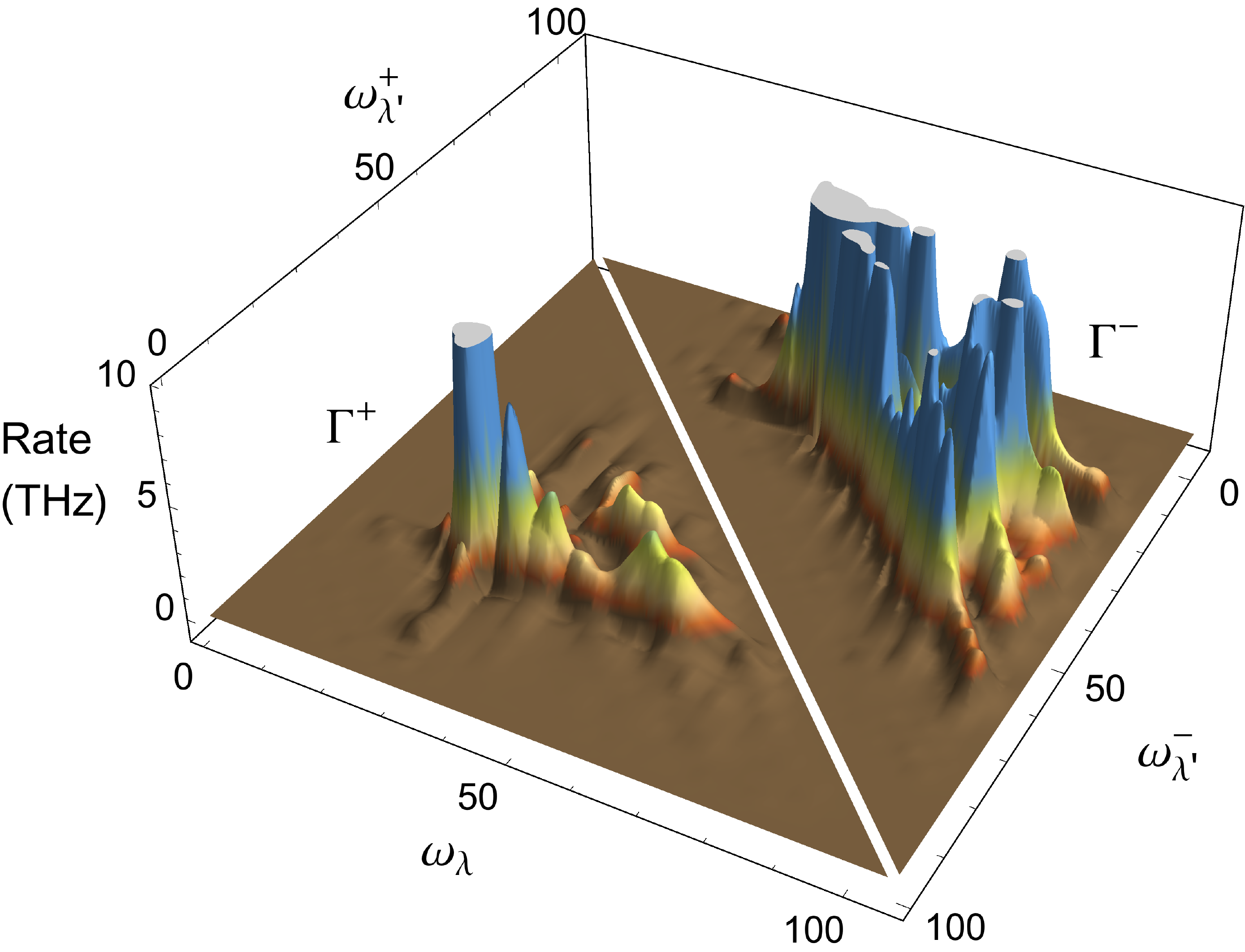}
	\caption{(color online) Scattering rates $\Gamma^+$ and  $\Gamma^-$ for Ba$_2$AuBi at 300~K. The left segment shows the absorption rates, whereas the right panel indicates the rates for emission processes. The corresponding phonon frequencies $\omega$ are given in units of cm$^{-1}$.}
	\label{fig:scatterrate}
\end{figure}

The mode dependent phonon lifetime $\tau$  shows a dip at $\omega\approx35$~cm$^{-1}$, another indication of strong rattling according to the resonant scattering model~\cite{toberer_phonon_2011} 
(see Fig.~S3).
The strong influence of the rattling modes on $\kappa_\text{L}$ can be seen in Fig. \ref{fig:scatterrate}, where the scattering rates $\Gamma^\pm$ for the 3-phonon interaction of both the absorption processes ($\Gamma^+$ for $\lambda+\lambda'\rightarrow\lambda''$) and emission processes  ($\Gamma^-$ for $\lambda\rightarrow\lambda'+\lambda''$) are shown~\cite{mingo_ab_2014}. Here we used  $\lambda\equiv(\alpha,\textbf{q})$ to describe phonons with branch index $\alpha$ and  wave vector $\textbf{q}$. As expected from the resonant scattering model, there is a strong contribution to $\Gamma^+$ due to rattling (left segment), but the rattlers also interact with the optical modes in the emission processes (right segment, $\Gamma^-$), which additionally impedes the transport of heat carried by optical phonons. Furthermore, there are two general chemical trends that can be observed in Fig.~\ref{fig:kappa}. First, the overall value of $\kappa_\text{L}$ decreases strongly with each period of the alkali earth metal elements $X_2$. This behavior can be simply attributed to the size of $X$ element, which, depending on its radius, creates larger or smaller voids for the $Y$ atoms, giving it more or less space to rattle. Second, and to a lesser extent, $\kappa_\text{L}$ decreases with the overall phonon frequencies at a given element $X$ 
(see Fig.~S2),
which is a direct consequence of the mass of the $Z$ element.

In summary, we predict a new class of thermodynamically stable FH semiconductors with excellent thermoelectric properties from first principles calculations. The electronic transport properties are comparable to the well-studied thermoelectric FH compound Fe$_2$VAl with high PFs due to ``flat-and-dispersive'' valence bands.
Unlike known FH and HH compounds, our novel FH compounds have one order of magnititude lower lattice thermal conductivities, comparable to the value of SnSe.
The heavy noble metals Au and Hg are only weakly bonded such that they can freely rattle in the pseudo-cages composed of the nearest and next-nearest neighbors. Consequently, the strong scattering of the heat carrying acoustic phonons leads to a suppression of the thermal conductivity to values comparable to glass, which could be further reduced through structural engineering. In strong contrast to the common perception that only complex or host-guest structures with large unit cells can lead to such low thermal conductivity~\cite{toberer_phonon_2011}, R-Heuslers demonstrate that ultralow thermal conductivity can be realized in very simple crystal structures. These properties not only establish the novel R-Heuslers as very promising thermoelectric materials, but will moreover lead to fundamental advances in the design of novel materials for thermoelectric applications.


J.~H. and C.~W. acknowledge support via ONR STTR N00014-13-P-1056.
M.~A. gratefully acknowledges support from the Novartis Universit\"{a}t Basel Excellence Scholarship for Life Sciences and the Swiss National Science Foundation.
Y.~X., S.~S.~N., and V.~O. acknowledge support by the U.S. Department of Energy, Office of Science, Basic Energy Sciences, under Grant DEFG02-07ER46433.
V.~I.~H. acknowledges support from NSF grant DMR-1309957.
S.~H. acknowledges support by the U.S. Department of Energy, Office of Science, Office of Basic Energy Sciences, under Award Number DE-SC0014520.
Computational resources from the Swiss National Supercomputing Center (CSCS) in Lugano (projects s499 and s621) and the National Energy Research Scientific Computing Center, which is supported by the Office of Science of the U.S. Department of Energy under Contract No. DE-AC02-05CH11231, are acknowledged.


\begin{thebibliography}{56}%
\makeatletter
\providecommand \@ifxundefined [1]{%
 \@ifx{#1\undefined}
}%
\providecommand \@ifnum [1]{%
 \ifnum #1\expandafter \@firstoftwo
 \else \expandafter \@secondoftwo
 \fi
}%
\providecommand \@ifx [1]{%
 \ifx #1\expandafter \@firstoftwo
 \else \expandafter \@secondoftwo
 \fi
}%
\providecommand \natexlab [1]{#1}%
\providecommand \enquote  [1]{``#1''}%
\providecommand \bibnamefont  [1]{#1}%
\providecommand \bibfnamefont [1]{#1}%
\providecommand \citenamefont [1]{#1}%
\providecommand \href@noop [0]{\@secondoftwo}%
\providecommand \href [0]{\begingroup \@sanitize@url \@href}%
\providecommand \@href[1]{\@@startlink{#1}\@@href}%
\providecommand \@@href[1]{\endgroup#1\@@endlink}%
\providecommand \@sanitize@url [0]{\catcode `\\12\catcode `\$12\catcode
  `\&12\catcode `\#12\catcode `\^12\catcode `\_12\catcode `\%12\relax}%
\providecommand \@@startlink[1]{}%
\providecommand \@@endlink[0]{}%
\providecommand \url  [0]{\begingroup\@sanitize@url \@url }%
\providecommand \@url [1]{\endgroup\@href {#1}{\urlprefix }}%
\providecommand \urlprefix  [0]{URL }%
\providecommand \Eprint [0]{\href }%
\providecommand \doibase [0]{http://dx.doi.org/}%
\providecommand \selectlanguage [0]{\@gobble}%
\providecommand \bibinfo  [0]{\@secondoftwo}%
\providecommand \bibfield  [0]{\@secondoftwo}%
\providecommand \translation [1]{[#1]}%
\providecommand \BibitemOpen [0]{}%
\providecommand \bibitemStop [0]{}%
\providecommand \bibitemNoStop [0]{.\EOS\space}%
\providecommand \EOS [0]{\spacefactor3000\relax}%
\providecommand \BibitemShut  [1]{\csname bibitem#1\endcsname}%
\let\auto@bib@innerbib\@empty
\bibitem [{\citenamefont {Snyder}\ and\ \citenamefont
  {Toberer}(2008)}]{snyder_complex_2008}%
  \BibitemOpen
  \bibfield  {author} {\bibinfo {author} {\bibfnamefont {G.~J.}\ \bibnamefont
  {Snyder}}\ and\ \bibinfo {author} {\bibfnamefont {E.~S.}\ \bibnamefont
  {Toberer}},\ }\href {\doibase 10.1038/nmat2090} {\bibfield  {journal}
  {\bibinfo  {journal} {Nat. Mater.}\ }\textbf {\bibinfo {volume} {7}},\
  \bibinfo {pages} {105} (\bibinfo {year} {2008})}\BibitemShut {NoStop}%
\bibitem [{\citenamefont {Rowe}(2005)}]{rowe2005thermoelectrics}%
  \BibitemOpen
  \bibfield  {author} {\bibinfo {author} {\bibfnamefont {D.}~\bibnamefont
  {Rowe}},\ }\href {http://books.google.gr/books?id=0iwERQe5IKQC} {\emph
  {\bibinfo {title} {Thermoelectrics Handbook: Macro to Nano}}}\ (\bibinfo
  {publisher} {CRC Press},\ \bibinfo {year} {2005})\BibitemShut {NoStop}%
\bibitem [{\citenamefont {Wright}(1958)}]{wright_thermoelectric_1958}%
  \BibitemOpen
  \bibfield  {author} {\bibinfo {author} {\bibfnamefont {D.~A.}\ \bibnamefont
  {Wright}},\ }\href {\doibase 10.1038/181834a0} {\bibfield  {journal}
  {\bibinfo  {journal} {Nature}\ }\textbf {\bibinfo {volume} {181}},\ \bibinfo
  {pages} {834} (\bibinfo {year} {1958})}\BibitemShut {NoStop}%
\bibitem [{\citenamefont {Snyder}\ \emph {et~al.}(2004)\citenamefont {Snyder},
  \citenamefont {Christensen}, \citenamefont {Nishikori}, \citenamefont
  {Caillat},\ and\ \citenamefont {B.}}]{snyder2004}%
  \BibitemOpen
  \bibfield  {author} {\bibinfo {author} {\bibfnamefont {G.~J.}\ \bibnamefont
  {Snyder}}, \bibinfo {author} {\bibfnamefont {M.}~\bibnamefont {Christensen}},
  \bibinfo {author} {\bibfnamefont {E.}~\bibnamefont {Nishikori}}, \bibinfo
  {author} {\bibfnamefont {T.}~\bibnamefont {Caillat}}, \ and\ \bibinfo
  {author} {\bibfnamefont {I.~B.}\ \bibnamefont {B.}},\ }\href {\doibase
  10.1038/nmat1154} {\bibfield  {journal} {\bibinfo  {journal} {Nat. Mater.}\
  }\textbf {\bibinfo {volume} {3}},\ \bibinfo {pages} {458} (\bibinfo {year}
  {2004})}\BibitemShut {NoStop}%
\bibitem [{\citenamefont {Bjerg}, \citenamefont {Madsen},\ and\ \citenamefont
  {Iversen}(2011)}]{bjerg_enhanced_2011}%
  \BibitemOpen
  \bibfield  {author} {\bibinfo {author} {\bibfnamefont {L.}~\bibnamefont
  {Bjerg}}, \bibinfo {author} {\bibfnamefont {G.~K.~H.}\ \bibnamefont
  {Madsen}}, \ and\ \bibinfo {author} {\bibfnamefont {B.~B.}\ \bibnamefont
  {Iversen}},\ }\href {\doibase 10.1021/cm201271d} {\bibfield  {journal}
  {\bibinfo  {journal} {Chem. Mater.}\ }\textbf {\bibinfo {volume} {23}},\
  \bibinfo {pages} {3907} (\bibinfo {year} {2011})}\BibitemShut {NoStop}%
\bibitem [{\citenamefont {Bjerg}, \citenamefont {Iversen},\ and\ \citenamefont
  {Madsen}(2014)}]{bjerg_modeling_2014}%
  \BibitemOpen
  \bibfield  {author} {\bibinfo {author} {\bibfnamefont {L.}~\bibnamefont
  {Bjerg}}, \bibinfo {author} {\bibfnamefont {B.~B.}\ \bibnamefont {Iversen}},
  \ and\ \bibinfo {author} {\bibfnamefont {G.~K.~H.}\ \bibnamefont {Madsen}},\
  }\href {\doibase 10.1103/PhysRevB.89.024304} {\bibfield  {journal} {\bibinfo
  {journal} {Phys. Rev. B}\ }\textbf {\bibinfo {volume} {89}},\ \bibinfo
  {pages} {024304} (\bibinfo {year} {2014})}\BibitemShut {NoStop}%
\bibitem [{\citenamefont {Tadano}, \citenamefont {Gohda},\ and\ \citenamefont
  {Tsuneyuki}(2015)}]{tadano_impact_2015}%
  \BibitemOpen
  \bibfield  {author} {\bibinfo {author} {\bibfnamefont {T.}~\bibnamefont
  {Tadano}}, \bibinfo {author} {\bibfnamefont {Y.}~\bibnamefont {Gohda}}, \
  and\ \bibinfo {author} {\bibfnamefont {S.}~\bibnamefont {Tsuneyuki}},\ }\href
  {\doibase 10.1103/PhysRevLett.114.095501} {\bibfield  {journal} {\bibinfo
  {journal} {Phys. Rev. Lett.}\ }\textbf {\bibinfo {volume} {114}},\ \bibinfo
  {pages} {095501} (\bibinfo {year} {2015})}\BibitemShut {NoStop}%
\bibitem [{\citenamefont {Wu}\ \emph {et~al.}(2014)\citenamefont {Wu},
  \citenamefont {Zhao}, \citenamefont {Zheng}, \citenamefont {Wu},
  \citenamefont {Pei}, \citenamefont {Tong}, \citenamefont {Kanatzidis},\ and\
  \citenamefont {He}}]{wu_broad_2014}%
  \BibitemOpen
  \bibfield  {author} {\bibinfo {author} {\bibfnamefont {H.~J.}\ \bibnamefont
  {Wu}}, \bibinfo {author} {\bibfnamefont {L.-D.}\ \bibnamefont {Zhao}},
  \bibinfo {author} {\bibfnamefont {F.~S.}\ \bibnamefont {Zheng}}, \bibinfo
  {author} {\bibfnamefont {D.}~\bibnamefont {Wu}}, \bibinfo {author}
  {\bibfnamefont {Y.~L.}\ \bibnamefont {Pei}}, \bibinfo {author} {\bibfnamefont
  {X.}~\bibnamefont {Tong}}, \bibinfo {author} {\bibfnamefont {M.~G.}\
  \bibnamefont {Kanatzidis}}, \ and\ \bibinfo {author} {\bibfnamefont {J.~Q.}\
  \bibnamefont {He}},\ }\href {\doibase 10.1038/ncomms5515} {\bibfield
  {journal} {\bibinfo  {journal} {Nat. Commun.}\ }\textbf {\bibinfo {volume}
  {5}},\ \bibinfo {pages} {4515} (\bibinfo {year} {2014})}\BibitemShut
  {NoStop}%
\bibitem [{\citenamefont {Uher}(2001)}]{uher_chapter_2001}%
  \BibitemOpen
  \bibfield  {author} {\bibinfo {author} {\bibfnamefont {C.}~\bibnamefont
  {Uher}},\ }in\ \href
  {http://www.sciencedirect.com/science/article/pii/S0080878401801514} {\emph
  {\bibinfo {booktitle} {Semiconductors and {Semimetals}}}},\ \bibinfo {series}
  {Recent {Trends} in {Thermoelectric} {Materials} {Research} {I}},
  Vol.~\bibinfo {volume} {69},\ \bibinfo {editor} {edited by\ \bibinfo {editor}
  {\bibfnamefont {T.~M.}\ \bibnamefont {Tritt}}}\ (\bibinfo  {publisher}
  {Elsevier},\ \bibinfo {year} {2001})\ pp.\ \bibinfo {pages}
  {139--253}\BibitemShut {NoStop}%
\bibitem [{\citenamefont {Nolas}, \citenamefont {Slack},\ and\ \citenamefont
  {Schujman}(2001)}]{nolas_chapter_2001}%
  \BibitemOpen
  \bibfield  {author} {\bibinfo {author} {\bibfnamefont {G.~S.}\ \bibnamefont
  {Nolas}}, \bibinfo {author} {\bibfnamefont {G.~A.}\ \bibnamefont {Slack}}, \
  and\ \bibinfo {author} {\bibfnamefont {S.~B.}\ \bibnamefont {Schujman}},\
  }in\ \href
  {http://www.sciencedirect.com/science/article/pii/S0080878401801526} {\emph
  {\bibinfo {booktitle} {Semiconductors and {Semimetals}}}},\ \bibinfo {series}
  {Recent {Trends} in {Thermoelectric} {Materials} {Research} {I}},
  Vol.~\bibinfo {volume} {69},\ \bibinfo {editor} {edited by\ \bibinfo {editor}
  {\bibfnamefont {T.~M.}\ \bibnamefont {Tritt}}}\ (\bibinfo  {publisher}
  {Elsevier},\ \bibinfo {year} {2001})\ pp.\ \bibinfo {pages}
  {255--300}\BibitemShut {NoStop}%
\bibitem [{\citenamefont {Nolas}, \citenamefont {Poon},\ and\ \citenamefont
  {Kanatzidis}(2006)}]{nolas_recent_2006}%
  \BibitemOpen
  \bibfield  {author} {\bibinfo {author} {\bibfnamefont {G.~S.}\ \bibnamefont
  {Nolas}}, \bibinfo {author} {\bibfnamefont {J.}~\bibnamefont {Poon}}, \ and\
  \bibinfo {author} {\bibfnamefont {M.}~\bibnamefont {Kanatzidis}},\ }\href
  {\doibase 10.1557/mrs2006.45} {\bibfield  {journal} {\bibinfo  {journal} {MRS
  Bulletin}\ }\textbf {\bibinfo {volume} {31}},\ \bibinfo {pages} {199}
  (\bibinfo {year} {2006})}\BibitemShut {NoStop}%
\bibitem [{\citenamefont {Lu}\ \emph {et~al.}(2013)\citenamefont {Lu},
  \citenamefont {Morelli}, \citenamefont {Xia}, \citenamefont {Zhou},
  \citenamefont {Ozoli\c{n}\v{s}}, \citenamefont {Chi}, \citenamefont {Zhou},\
  and\ \citenamefont {Uher}}]{Lu2013}%
  \BibitemOpen
  \bibfield  {author} {\bibinfo {author} {\bibfnamefont {X.}~\bibnamefont
  {Lu}}, \bibinfo {author} {\bibfnamefont {D.~T.}\ \bibnamefont {Morelli}},
  \bibinfo {author} {\bibfnamefont {Y.}~\bibnamefont {Xia}}, \bibinfo {author}
  {\bibfnamefont {F.}~\bibnamefont {Zhou}}, \bibinfo {author} {\bibfnamefont
  {V.}~\bibnamefont {Ozoli\c{n}\v{s}}}, \bibinfo {author} {\bibfnamefont
  {H.}~\bibnamefont {Chi}}, \bibinfo {author} {\bibfnamefont {X.}~\bibnamefont
  {Zhou}}, \ and\ \bibinfo {author} {\bibfnamefont {C.}~\bibnamefont {Uher}},\
  }\href {\doibase 10.1002/aenm.201200650} {\bibfield  {journal} {\bibinfo
  {journal} {Adv. Energy Mater.}\ }\textbf {\bibinfo {volume} {3}},\ \bibinfo
  {pages} {342} (\bibinfo {year} {2013})}\BibitemShut {NoStop}%
\bibitem [{\citenamefont {Zhao}\ \emph {et~al.}(2014)\citenamefont {Zhao},
  \citenamefont {Lo}, \citenamefont {Zhang}, \citenamefont {Sun}, \citenamefont
  {Tan}, \citenamefont {Uher}, \citenamefont {Wolverton}, \citenamefont
  {Dravid},\ and\ \citenamefont {Kanatzidis}}]{SnSe}%
  \BibitemOpen
  \bibfield  {author} {\bibinfo {author} {\bibfnamefont {L.-D.}\ \bibnamefont
  {Zhao}}, \bibinfo {author} {\bibfnamefont {S.-H.}\ \bibnamefont {Lo}},
  \bibinfo {author} {\bibfnamefont {Y.}~\bibnamefont {Zhang}}, \bibinfo
  {author} {\bibfnamefont {H.}~\bibnamefont {Sun}}, \bibinfo {author}
  {\bibfnamefont {G.}~\bibnamefont {Tan}}, \bibinfo {author} {\bibfnamefont
  {C.}~\bibnamefont {Uher}}, \bibinfo {author} {\bibfnamefont {C.}~\bibnamefont
  {Wolverton}}, \bibinfo {author} {\bibfnamefont {V.~P.}\ \bibnamefont
  {Dravid}}, \ and\ \bibinfo {author} {\bibfnamefont {M.~G.}\ \bibnamefont
  {Kanatzidis}},\ }\href {\doibase 10.1038/nature13184} {\bibfield  {journal}
  {\bibinfo  {journal} {Nature}\ }\textbf {\bibinfo {volume} {508}},\ \bibinfo
  {pages} {373} (\bibinfo {year} {2014})}\BibitemShut {NoStop}%
\bibitem [{\citenamefont {Li}\ \emph {et~al.}(2015)\citenamefont {Li},
  \citenamefont {Hong}, \citenamefont {May}, \citenamefont {Bansal},
  \citenamefont {Chi}, \citenamefont {Hong}, \citenamefont {Ehlers},\ and\
  \citenamefont {Delaire}}]{doi:10.1038/nphys3492}%
  \BibitemOpen
  \bibfield  {author} {\bibinfo {author} {\bibfnamefont {C.~W.}\ \bibnamefont
  {Li}}, \bibinfo {author} {\bibfnamefont {J.}~\bibnamefont {Hong}}, \bibinfo
  {author} {\bibfnamefont {A.~F.}\ \bibnamefont {May}}, \bibinfo {author}
  {\bibfnamefont {D.}~\bibnamefont {Bansal}}, \bibinfo {author} {\bibfnamefont
  {S.}~\bibnamefont {Chi}}, \bibinfo {author} {\bibfnamefont {T.}~\bibnamefont
  {Hong}}, \bibinfo {author} {\bibfnamefont {G.}~\bibnamefont {Ehlers}}, \ and\
  \bibinfo {author} {\bibfnamefont {O.}~\bibnamefont {Delaire}},\ }\href
  {\doibase 10.1038/nphys3492} {\bibfield  {journal} {\bibinfo  {journal} {Nat.
  Phys.}\ }\textbf {\bibinfo {volume} {11}},\ \bibinfo {pages} {1063–1069}
  (\bibinfo {year} {2015})}\BibitemShut {NoStop}%
\bibitem [{\citenamefont {Zhu}\ \emph {et~al.}(2015)\citenamefont {Zhu},
  \citenamefont {Fu}, \citenamefont {Xie}, \citenamefont {Liu},\ and\
  \citenamefont {Zhao}}]{zhu_high_2015}%
  \BibitemOpen
  \bibfield  {author} {\bibinfo {author} {\bibfnamefont {T.}~\bibnamefont
  {Zhu}}, \bibinfo {author} {\bibfnamefont {C.}~\bibnamefont {Fu}}, \bibinfo
  {author} {\bibfnamefont {H.}~\bibnamefont {Xie}}, \bibinfo {author}
  {\bibfnamefont {Y.}~\bibnamefont {Liu}}, \ and\ \bibinfo {author}
  {\bibfnamefont {X.}~\bibnamefont {Zhao}},\ }\href {\doibase
  10.1002/aenm.201500588} {\bibfield  {journal} {\bibinfo  {journal} {Adv.
  Energy Mater.}\ }\textbf {\bibinfo {volume} {5}},\ \bibinfo {pages} {1500588}
  (\bibinfo {year} {2015})}\BibitemShut {NoStop}%
\bibitem [{\citenamefont {Rao}, \citenamefont {Ji},\ and\ \citenamefont
  {Tritt}(2006)}]{rao_properties_2006}%
  \BibitemOpen
  \bibfield  {author} {\bibinfo {author} {\bibfnamefont {A.~M.}\ \bibnamefont
  {Rao}}, \bibinfo {author} {\bibfnamefont {X.}~\bibnamefont {Ji}}, \ and\
  \bibinfo {author} {\bibfnamefont {T.~M.}\ \bibnamefont {Tritt}},\ }\href
  {\doibase 10.1557/mrs2006.48} {\bibfield  {journal} {\bibinfo  {journal}
  {{MRS} Bulletin}\ }\textbf {\bibinfo {volume} {31}},\ \bibinfo {pages} {218}
  (\bibinfo {year} {2006})}\BibitemShut {NoStop}%
\bibitem [{\citenamefont {Chen}\ and\ \citenamefont {Ren}(2013)}]{Chen2013387}%
  \BibitemOpen
  \bibfield  {author} {\bibinfo {author} {\bibfnamefont {S.}~\bibnamefont
  {Chen}}\ and\ \bibinfo {author} {\bibfnamefont {Z.}~\bibnamefont {Ren}},\
  }\href {\doibase http://dx.doi.org/10.1016/j.mattod.2013.09.015} {\bibfield
  {journal} {\bibinfo  {journal} {Materials Today}\ }\textbf {\bibinfo {volume}
  {16}},\ \bibinfo {pages} {387 } (\bibinfo {year} {2013})}\BibitemShut
  {NoStop}%
\bibitem [{\citenamefont {Carrete}\ \emph {et~al.}(2014)\citenamefont
  {Carrete}, \citenamefont {Li}, \citenamefont {Mingo}, \citenamefont {Wang},\
  and\ \citenamefont {Curtarolo}}]{PhysRevX.4.011019}%
  \BibitemOpen
  \bibfield  {author} {\bibinfo {author} {\bibfnamefont {J.}~\bibnamefont
  {Carrete}}, \bibinfo {author} {\bibfnamefont {W.}~\bibnamefont {Li}},
  \bibinfo {author} {\bibfnamefont {N.}~\bibnamefont {Mingo}}, \bibinfo
  {author} {\bibfnamefont {S.}~\bibnamefont {Wang}}, \ and\ \bibinfo {author}
  {\bibfnamefont {S.}~\bibnamefont {Curtarolo}},\ }\href {\doibase
  10.1103/PhysRevX.4.011019} {\bibfield  {journal} {\bibinfo  {journal} {Phys.
  Rev. X}\ }\textbf {\bibinfo {volume} {4}},\ \bibinfo {pages} {011019}
  (\bibinfo {year} {2014})}\BibitemShut {NoStop}%
\bibitem [{\citenamefont {Heusler}(1903)}]{Heusler}%
  \BibitemOpen
  \bibfield  {author} {\bibinfo {author} {\bibfnamefont {F.}~\bibnamefont
  {Heusler}},\ }\href@noop {} {\bibfield  {journal} {\bibinfo  {journal}
  {Deutsche Physikalische Gesellschaft}\ }\textbf {\bibinfo {volume} {5}},\
  \bibinfo {pages} {219} (\bibinfo {year} {1903})}\BibitemShut {NoStop}%
\bibitem [{\citenamefont {Galanakis}, \citenamefont {Dederichs},\ and\
  \citenamefont {Papanikolaou}(2002)}]{PhysRevB.66.174429}%
  \BibitemOpen
  \bibfield  {author} {\bibinfo {author} {\bibfnamefont {I.}~\bibnamefont
  {Galanakis}}, \bibinfo {author} {\bibfnamefont {P.~H.}\ \bibnamefont
  {Dederichs}}, \ and\ \bibinfo {author} {\bibfnamefont {N.}~\bibnamefont
  {Papanikolaou}},\ }\href {\doibase 10.1103/PhysRevB.66.174429} {\bibfield
  {journal} {\bibinfo  {journal} {Phys. Rev. B}\ }\textbf {\bibinfo {volume}
  {66}},\ \bibinfo {pages} {174429} (\bibinfo {year} {2002})}\BibitemShut
  {NoStop}%
\bibitem [{\citenamefont {Nishino}, \citenamefont {Deguchi},\ and\
  \citenamefont {Mizutani}(2006)}]{nishino_thermal_2006}%
  \BibitemOpen
  \bibfield  {author} {\bibinfo {author} {\bibfnamefont {Y.}~\bibnamefont
  {Nishino}}, \bibinfo {author} {\bibfnamefont {S.}~\bibnamefont {Deguchi}}, \
  and\ \bibinfo {author} {\bibfnamefont {U.}~\bibnamefont {Mizutani}},\ }\href
  {\doibase 10.1103/PhysRevB.74.115115} {\bibfield  {journal} {\bibinfo
  {journal} {Phys. Rev. B}\ }\textbf {\bibinfo {volume} {74}},\ \bibinfo
  {pages} {115115} (\bibinfo {year} {2006})}\BibitemShut {NoStop}%
\bibitem [{\citenamefont {Vasundhara}, \citenamefont {Srinivas},\ and\
  \citenamefont {Rao}(2008)}]{vasundhara_electronic_2008}%
  \BibitemOpen
  \bibfield  {author} {\bibinfo {author} {\bibfnamefont {M.}~\bibnamefont
  {Vasundhara}}, \bibinfo {author} {\bibfnamefont {V.}~\bibnamefont
  {Srinivas}}, \ and\ \bibinfo {author} {\bibfnamefont {V.~V.}\ \bibnamefont
  {Rao}},\ }\href {\doibase 10.1103/PhysRevB.77.224415} {\bibfield  {journal}
  {\bibinfo  {journal} {Phys. Rev. B}\ }\textbf {\bibinfo {volume} {77}},\
  \bibinfo {pages} {224415} (\bibinfo {year} {2008})}\BibitemShut {NoStop}%
\bibitem [{\citenamefont {Skoug}\ \emph {et~al.}(2009)\citenamefont {Skoug},
  \citenamefont {Zhou}, \citenamefont {Pei},\ and\ \citenamefont
  {Morelli}}]{skoug_high_2009}%
  \BibitemOpen
  \bibfield  {author} {\bibinfo {author} {\bibfnamefont {E.~J.}\ \bibnamefont
  {Skoug}}, \bibinfo {author} {\bibfnamefont {C.}~\bibnamefont {Zhou}},
  \bibinfo {author} {\bibfnamefont {Y.}~\bibnamefont {Pei}}, \ and\ \bibinfo
  {author} {\bibfnamefont {D.~T.}\ \bibnamefont {Morelli}},\ }\href {\doibase
  10.1007/s11664-008-0626-x} {\bibfield  {journal} {\bibinfo  {journal} {J.
  Electron. Mater.}\ }\textbf {\bibinfo {volume} {38}},\ \bibinfo {pages}
  {1221} (\bibinfo {year} {2009})}\BibitemShut {NoStop}%
\bibitem [{\citenamefont {Terazawa}\ \emph {et~al.}(2011)\citenamefont
  {Terazawa}, \citenamefont {Mikami}, \citenamefont {Itoh},\ and\ \citenamefont
  {Takeuchi}}]{terazawa_effects_2011}%
  \BibitemOpen
  \bibfield  {author} {\bibinfo {author} {\bibfnamefont {Y.}~\bibnamefont
  {Terazawa}}, \bibinfo {author} {\bibfnamefont {M.}~\bibnamefont {Mikami}},
  \bibinfo {author} {\bibfnamefont {T.}~\bibnamefont {Itoh}}, \ and\ \bibinfo
  {author} {\bibfnamefont {T.}~\bibnamefont {Takeuchi}},\ }\href {\doibase
  10.1007/s11664-011-1862-z} {\bibfield  {journal} {\bibinfo  {journal} {J.
  Electron. Mater.}\ }\textbf {\bibinfo {volume} {41}},\ \bibinfo {pages}
  {1348} (\bibinfo {year} {2011})}\BibitemShut {NoStop}%
\bibitem [{\citenamefont {Mikami}\ \emph {et~al.}(2012)\citenamefont {Mikami},
  \citenamefont {Kinemuchi}, \citenamefont {Ozaki}, \citenamefont {Terazawa},\
  and\ \citenamefont {Takeuchi}}]{mikami_thermoelectric_2012}%
  \BibitemOpen
  \bibfield  {author} {\bibinfo {author} {\bibfnamefont {M.}~\bibnamefont
  {Mikami}}, \bibinfo {author} {\bibfnamefont {Y.}~\bibnamefont {Kinemuchi}},
  \bibinfo {author} {\bibfnamefont {K.}~\bibnamefont {Ozaki}}, \bibinfo
  {author} {\bibfnamefont {Y.}~\bibnamefont {Terazawa}}, \ and\ \bibinfo
  {author} {\bibfnamefont {T.}~\bibnamefont {Takeuchi}},\ }\href {\doibase
  10.1063/1.4710990} {\bibfield  {journal} {\bibinfo  {journal} {J. Appl.
  Phys.}\ }\textbf {\bibinfo {volume} {111}},\ \bibinfo {pages} {093710}
  (\bibinfo {year} {2012})}\BibitemShut {NoStop}%
\bibitem [{\citenamefont {Bilc}\ \emph {et~al.}(2015)\citenamefont {Bilc},
  \citenamefont {Hautier}, \citenamefont {Waroquiers}, \citenamefont
  {Rignanese},\ and\ \citenamefont {Ghosez}}]{bilc_low-dimensional_2015}%
  \BibitemOpen
  \bibfield  {author} {\bibinfo {author} {\bibfnamefont {D.~I.}\ \bibnamefont
  {Bilc}}, \bibinfo {author} {\bibfnamefont {G.}~\bibnamefont {Hautier}},
  \bibinfo {author} {\bibfnamefont {D.}~\bibnamefont {Waroquiers}}, \bibinfo
  {author} {\bibfnamefont {G.-M.}\ \bibnamefont {Rignanese}}, \ and\ \bibinfo
  {author} {\bibfnamefont {P.}~\bibnamefont {Ghosez}},\ }\href {\doibase
  10.1103/PhysRevLett.114.136601} {\bibfield  {journal} {\bibinfo  {journal}
  {Phys. Rev. Lett.}\ }\textbf {\bibinfo {volume} {114}},\ \bibinfo {pages}
  {136601} (\bibinfo {year} {2015})}\BibitemShut {NoStop}%
\bibitem [{\citenamefont {Graf}, \citenamefont {Felser},\ and\ \citenamefont
  {Parkin}(2011)}]{Felser11}%
  \BibitemOpen
  \bibfield  {author} {\bibinfo {author} {\bibfnamefont {T.}~\bibnamefont
  {Graf}}, \bibinfo {author} {\bibfnamefont {C.}~\bibnamefont {Felser}}, \ and\
  \bibinfo {author} {\bibfnamefont {S.~S.}\ \bibnamefont {Parkin}},\ }\href
  {\doibase http://dx.doi.org/10.1016/j.progsolidstchem.2011.02.001} {\bibfield
   {journal} {\bibinfo  {journal} {Prog. Solid State Chem.}\ }\textbf {\bibinfo
  {volume} {39}},\ \bibinfo {pages} {1 } (\bibinfo {year} {2011})}\BibitemShut
  {NoStop}%
\bibitem [{\citenamefont {Felser}\ and\ \citenamefont
  {Hirohata}(2015)}]{felser_heusler_2015}%
  \BibitemOpen
  \bibfield  {author} {\bibinfo {author} {\bibfnamefont {C.}~\bibnamefont
  {Felser}}\ and\ \bibinfo {author} {\bibfnamefont {A.}~\bibnamefont
  {Hirohata}},\ }\href {http://www.springer.com/us/book/9783319214481} {\emph
  {\bibinfo {title} {Heusler {Alloys}: {Properties}, {Growth},
  {Applications}}}}\ (\bibinfo  {publisher} {Springer},\ \bibinfo {year}
  {2015})\BibitemShut {NoStop}%
\bibitem [{qmp()}]{qmpy}%
  \BibitemOpen
  \href@noop {} {\enquote {\bibinfo {title} {Qmpy (suite of computational
  materials science tools)},}\ }\bibinfo {howpublished}
  {\url{https://pypi.python.org/pypi/qmpy}}\BibitemShut {NoStop}%
\bibitem [{sup()}]{suppl}%
  \BibitemOpen
  \href@noop {} {\enquote {\bibinfo {title} {{Supplementary materials: Contains
  details on the computational methods employed in this work. Furthermore,
  detailed descriptions of the thermodynamic stabilities, prototype crystal
  structures, phonon dispersion, phonon lifetimes and mean free paths,
  electronic band structure, and power factors are presented. Tables S1-S4 and
  Figures S1-S6.}}}\ }\BibitemShut {NoStop}%
\bibitem [{\citenamefont {Bl\"{o}chl}(1994)}]{PAW-Blochl-1994}%
  \BibitemOpen
  \bibfield  {author} {\bibinfo {author} {\bibfnamefont {P.~E.}\ \bibnamefont
  {Bl\"{o}chl}},\ }\href {\doibase 10.1103/PhysRevB.50.17953} {\bibfield
  {journal} {\bibinfo  {journal} {Phys. Rev. B}\ }\textbf {\bibinfo {volume}
  {50}},\ \bibinfo {pages} {17953} (\bibinfo {year} {1994})}\BibitemShut
  {NoStop}%
\bibitem [{\citenamefont {Kresse}\ and\ \citenamefont
  {Furthm\"{u}ller}(1996)}]{VASP-Kresse-1996}%
  \BibitemOpen
  \bibfield  {author} {\bibinfo {author} {\bibfnamefont {G.}~\bibnamefont
  {Kresse}}\ and\ \bibinfo {author} {\bibfnamefont {J.}~\bibnamefont
  {Furthm\"{u}ller}},\ }\href {\doibase 10.1016/0927-0256(96)00008-0}
  {\bibfield  {journal} {\bibinfo  {journal} {Comput. Mater. Sci.}\ }\textbf
  {\bibinfo {volume} {6}},\ \bibinfo {pages} {15} (\bibinfo {year}
  {1996})}\BibitemShut {NoStop}%
\bibitem [{\citenamefont {Kresse}\ and\ \citenamefont
  {Joubert}(1999)}]{VASP-Kresse-1999}%
  \BibitemOpen
  \bibfield  {author} {\bibinfo {author} {\bibfnamefont {G.}~\bibnamefont
  {Kresse}}\ and\ \bibinfo {author} {\bibfnamefont {D.}~\bibnamefont
  {Joubert}},\ }\href {\doibase 10.1103/PhysRevB.59.1758} {\bibfield  {journal}
  {\bibinfo  {journal} {Phys. Rev. B}\ }\textbf {\bibinfo {volume} {59}},\
  \bibinfo {pages} {1758} (\bibinfo {year} {1999})}\BibitemShut {NoStop}%
\bibitem [{\citenamefont {Perdew}, \citenamefont {Burke},\ and\ \citenamefont
  {Ernzerhof}(1996)}]{Perdew-PBE-1996}%
  \BibitemOpen
  \bibfield  {author} {\bibinfo {author} {\bibfnamefont {J.~P.}\ \bibnamefont
  {Perdew}}, \bibinfo {author} {\bibfnamefont {K.}~\bibnamefont {Burke}}, \
  and\ \bibinfo {author} {\bibfnamefont {M.}~\bibnamefont {Ernzerhof}},\ }\href
  {\doibase doi.org/10.1103/PhysRevLett.77.3865} {\bibfield  {journal}
  {\bibinfo  {journal} {Phys. Rev. Lett.}\ }\textbf {\bibinfo {volume} {77}},\
  \bibinfo {pages} {3865} (\bibinfo {year} {1996})}\BibitemShut {NoStop}%
\bibitem [{\citenamefont {Saal}\ \emph {et~al.}(2013)\citenamefont {Saal},
  \citenamefont {Kirklin}, \citenamefont {Aykol}, \citenamefont {Meredig},\
  and\ \citenamefont {Wolverton}}]{OQMD}%
  \BibitemOpen
  \bibfield  {author} {\bibinfo {author} {\bibfnamefont {J.}~\bibnamefont
  {Saal}}, \bibinfo {author} {\bibfnamefont {S.}~\bibnamefont {Kirklin}},
  \bibinfo {author} {\bibfnamefont {M.}~\bibnamefont {Aykol}}, \bibinfo
  {author} {\bibfnamefont {B.}~\bibnamefont {Meredig}}, \ and\ \bibinfo
  {author} {\bibfnamefont {C.}~\bibnamefont {Wolverton}},\ }\href {\doibase
  10.1007/s11837-013-0755-4} {\bibfield  {journal} {\bibinfo  {journal} {JOM}\
  }\textbf {\bibinfo {volume} {65}},\ \bibinfo {pages} {1501} (\bibinfo {year}
  {2013})}\BibitemShut {NoStop}%
\bibitem [{\citenamefont {Kirklin}\ \emph {et~al.}(2015)\citenamefont
  {Kirklin}, \citenamefont {Saal}, \citenamefont {Meredig}, \citenamefont
  {Thompson}, \citenamefont {Doak}, \citenamefont {Aykol}, \citenamefont
  {Rühl},\ and\ \citenamefont {Wolverton}}]{OQMD2}%
  \BibitemOpen
  \bibfield  {author} {\bibinfo {author} {\bibfnamefont {S.}~\bibnamefont
  {Kirklin}}, \bibinfo {author} {\bibfnamefont {J.~E.}\ \bibnamefont {Saal}},
  \bibinfo {author} {\bibfnamefont {B.}~\bibnamefont {Meredig}}, \bibinfo
  {author} {\bibfnamefont {A.}~\bibnamefont {Thompson}}, \bibinfo {author}
  {\bibfnamefont {J.~W.}\ \bibnamefont {Doak}}, \bibinfo {author}
  {\bibfnamefont {M.}~\bibnamefont {Aykol}}, \bibinfo {author} {\bibfnamefont
  {S.}~\bibnamefont {Rühl}}, \ and\ \bibinfo {author} {\bibfnamefont
  {C.}~\bibnamefont {Wolverton}},\ }\href {\doibase
  10.1038/npjcompumats.2015.10} {\bibfield  {journal} {\bibinfo  {journal}
  {npj. Comput. Mater.}\ }\textbf {\bibinfo {volume} {1}},\ \bibinfo {pages}
  {15010} (\bibinfo {year} {2015})}\BibitemShut {NoStop}%
\bibitem [{\citenamefont {Goedecker}(2004)}]{mhm1}%
  \BibitemOpen
  \bibfield  {author} {\bibinfo {author} {\bibfnamefont {S.}~\bibnamefont
  {Goedecker}},\ }\href {\doibase 10.1063/1.1724816} {\bibfield  {journal}
  {\bibinfo  {journal} {J. Chem. Phys.}\ }\textbf {\bibinfo {volume} {120}},\
  \bibinfo {pages} {9911} (\bibinfo {year} {2004})}\BibitemShut {NoStop}%
\bibitem [{\citenamefont {Amsler}\ and\ \citenamefont
  {Goedecker}(2010)}]{mhm2}%
  \BibitemOpen
  \bibfield  {author} {\bibinfo {author} {\bibfnamefont {M.}~\bibnamefont
  {Amsler}}\ and\ \bibinfo {author} {\bibfnamefont {S.}~\bibnamefont
  {Goedecker}},\ }\href {\doibase 10.1063/1.3512900} {\bibfield  {journal}
  {\bibinfo  {journal} {J. Chem. Phys.}\ }\textbf {\bibinfo {volume} {133}},\
  \bibinfo {pages} {224104} (\bibinfo {year} {2010})}\BibitemShut {NoStop}%
\bibitem [{\citenamefont {Togo}, \citenamefont {Oba},\ and\ \citenamefont
  {Tanaka}(2008)}]{phonopy}%
  \BibitemOpen
  \bibfield  {author} {\bibinfo {author} {\bibfnamefont {A.}~\bibnamefont
  {Togo}}, \bibinfo {author} {\bibfnamefont {F.}~\bibnamefont {Oba}}, \ and\
  \bibinfo {author} {\bibfnamefont {I.}~\bibnamefont {Tanaka}},\ }\href
  {\doibase 10.1103/PhysRevB.78.134106} {\bibfield  {journal} {\bibinfo
  {journal} {Phys. Rev. B}\ }\textbf {\bibinfo {volume} {78}},\ \bibinfo
  {pages} {134106} (\bibinfo {year} {2008})}\BibitemShut {NoStop}%
\bibitem [{\citenamefont {Blaha}\ \emph {et~al.}(1990)\citenamefont {Blaha},
  \citenamefont {Schwarz}, \citenamefont {Sorantin},\ and\ \citenamefont
  {Trickey}}]{blaha_full-potential_1990}%
  \BibitemOpen
  \bibfield  {author} {\bibinfo {author} {\bibfnamefont {P.}~\bibnamefont
  {Blaha}}, \bibinfo {author} {\bibfnamefont {K.}~\bibnamefont {Schwarz}},
  \bibinfo {author} {\bibfnamefont {P.}~\bibnamefont {Sorantin}}, \ and\
  \bibinfo {author} {\bibfnamefont {S.~B.}\ \bibnamefont {Trickey}},\ }\href
  {\doibase 10.1016/0010-4655(90)90187-6} {\bibfield  {journal} {\bibinfo
  {journal} {Comput. Phys. Commun.}\ }\textbf {\bibinfo {volume} {59}},\
  \bibinfo {pages} {399} (\bibinfo {year} {1990})}\BibitemShut {NoStop}%
\bibitem [{\citenamefont {Heyd}, \citenamefont {Scuseria},\ and\ \citenamefont
  {Ernzerhof}(2006)}]{heyd_erratum:_2006}%
  \BibitemOpen
  \bibfield  {author} {\bibinfo {author} {\bibfnamefont {J.}~\bibnamefont
  {Heyd}}, \bibinfo {author} {\bibfnamefont {G.~E.}\ \bibnamefont {Scuseria}},
  \ and\ \bibinfo {author} {\bibfnamefont {M.}~\bibnamefont {Ernzerhof}},\
  }\href {\doibase doi.org/10.1063/1.2204597} {\bibfield  {journal} {\bibinfo
  {journal} {J. Chem. Phys.}\ }\textbf {\bibinfo {volume} {124}},\ \bibinfo
  {pages} {219906} (\bibinfo {year} {2006})}\BibitemShut {NoStop}%
\bibitem [{\citenamefont {Paier}\ \emph {et~al.}(2006)\citenamefont {Paier},
  \citenamefont {Marsman}, \citenamefont {Hummer}, \citenamefont {Kresse},
  \citenamefont {Gerber},\ and\ \citenamefont
  {\'{A}ngy\'{a}n}}]{paier_erratum:_2006}%
  \BibitemOpen
  \bibfield  {author} {\bibinfo {author} {\bibfnamefont {J.}~\bibnamefont
  {Paier}}, \bibinfo {author} {\bibfnamefont {M.}~\bibnamefont {Marsman}},
  \bibinfo {author} {\bibfnamefont {K.}~\bibnamefont {Hummer}}, \bibinfo
  {author} {\bibfnamefont {G.}~\bibnamefont {Kresse}}, \bibinfo {author}
  {\bibfnamefont {I.~C.}\ \bibnamefont {Gerber}}, \ and\ \bibinfo {author}
  {\bibfnamefont {J.~G.}\ \bibnamefont {\'{A}ngy\'{a}n}},\ }\href {\doibase
  doi.org/10.1063/1.2403866} {\bibfield  {journal} {\bibinfo  {journal} {J.
  Chem. Phys.}\ }\textbf {\bibinfo {volume} {125}},\ \bibinfo {pages} {249901}
  (\bibinfo {year} {2006})}\BibitemShut {NoStop}%
\bibitem [{\citenamefont {Heyd}, \citenamefont {Scuseria},\ and\ \citenamefont
  {Ernzerhof}(2003)}]{heyd_hybrid_2003}%
  \BibitemOpen
  \bibfield  {author} {\bibinfo {author} {\bibfnamefont {J.}~\bibnamefont
  {Heyd}}, \bibinfo {author} {\bibfnamefont {G.~E.}\ \bibnamefont {Scuseria}},
  \ and\ \bibinfo {author} {\bibfnamefont {M.}~\bibnamefont {Ernzerhof}},\
  }\href {\doibase doi.org/10.1063/1.1564060} {\bibfield  {journal} {\bibinfo
  {journal} {J. Chem. Phys.}\ }\textbf {\bibinfo {volume} {118}},\ \bibinfo
  {pages} {8207} (\bibinfo {year} {2003})}\BibitemShut {NoStop}%
\bibitem [{\citenamefont {Madsen}\ and\ \citenamefont
  {Singh}(2006)}]{Madsen200667}%
  \BibitemOpen
  \bibfield  {author} {\bibinfo {author} {\bibfnamefont {G.~K.}\ \bibnamefont
  {Madsen}}\ and\ \bibinfo {author} {\bibfnamefont {D.~J.}\ \bibnamefont
  {Singh}},\ }\href {\doibase http://dx.doi.org/10.1016/j.cpc.2006.03.007}
  {\bibfield  {journal} {\bibinfo  {journal} {Comput. Phys. Commun.}\ }\textbf
  {\bibinfo {volume} {175}},\ \bibinfo {pages} {67 } (\bibinfo {year}
  {2006})}\BibitemShut {NoStop}%
\bibitem [{\citenamefont {Zhou}\ \emph {et~al.}(2014)\citenamefont {Zhou},
  \citenamefont {Nielson}, \citenamefont {Xia},\ and\ \citenamefont
  {Ozoli\ifmmode \mbox{\c{n}}\else \c{n}\fi{}\ifmmode~\check{s}\else
  \v{s}\fi{}}}]{PhysRevLett.113.185501}%
  \BibitemOpen
  \bibfield  {author} {\bibinfo {author} {\bibfnamefont {F.}~\bibnamefont
  {Zhou}}, \bibinfo {author} {\bibfnamefont {W.}~\bibnamefont {Nielson}},
  \bibinfo {author} {\bibfnamefont {Y.}~\bibnamefont {Xia}}, \ and\ \bibinfo
  {author} {\bibfnamefont {V.}~\bibnamefont {Ozoli\ifmmode \mbox{\c{n}}\else
  \c{n}\fi{}\ifmmode~\check{s}\else \v{s}\fi{}}},\ }\href {\doibase
  10.1103/PhysRevLett.113.185501} {\bibfield  {journal} {\bibinfo  {journal}
  {Phys. Rev. Lett.}\ }\textbf {\bibinfo {volume} {113}},\ \bibinfo {pages}
  {185501} (\bibinfo {year} {2014})}\BibitemShut {NoStop}%
\bibitem [{\citenamefont {Li}\ \emph {et~al.}(2014)\citenamefont {Li},
  \citenamefont {Carrete}, \citenamefont {Katcho},\ and\ \citenamefont
  {Mingo}}]{ShengBTE_2014}%
  \BibitemOpen
  \bibfield  {author} {\bibinfo {author} {\bibfnamefont {W.}~\bibnamefont
  {Li}}, \bibinfo {author} {\bibfnamefont {J.}~\bibnamefont {Carrete}},
  \bibinfo {author} {\bibfnamefont {N.~A.}\ \bibnamefont {Katcho}}, \ and\
  \bibinfo {author} {\bibfnamefont {N.}~\bibnamefont {Mingo}},\ }\href
  {\doibase 10.1016/j.cpc.2014.02.015} {\bibfield  {journal} {\bibinfo
  {journal} {Comp. Phys. Commun.}\ }\textbf {\bibinfo {volume} {185}},\
  \bibinfo {pages} {1747–1758} (\bibinfo {year} {2014})}\BibitemShut
  {NoStop}%
\bibitem [{\citenamefont {Cahill}\ and\ \citenamefont
  {Pohl}(1988)}]{10.1146/annurev.pc.39.100188.000521}%
  \BibitemOpen
  \bibfield  {author} {\bibinfo {author} {\bibfnamefont {D.~G.}\ \bibnamefont
  {Cahill}}\ and\ \bibinfo {author} {\bibfnamefont {R.~O.}\ \bibnamefont
  {Pohl}},\ }\href {\doibase 10.1146/annurev.pc.39.100188.000521} {\bibfield
  {journal} {\bibinfo  {journal} {Annu. Rev. Phys. Chem.}\ }\textbf {\bibinfo
  {volume} {39}},\ \bibinfo {pages} {93} (\bibinfo {year} {1988})}\BibitemShut
  {NoStop}%
\bibitem [{\citenamefont {Toberer}, \citenamefont {Zevalkink},\ and\
  \citenamefont {Snyder}(2011)}]{toberer_phonon_2011}%
  \BibitemOpen
  \bibfield  {author} {\bibinfo {author} {\bibfnamefont {E.~S.}\ \bibnamefont
  {Toberer}}, \bibinfo {author} {\bibfnamefont {A.}~\bibnamefont {Zevalkink}},
  \ and\ \bibinfo {author} {\bibfnamefont {G.~J.}\ \bibnamefont {Snyder}},\
  }\href {\doibase 10.1039/C1JM11754H} {\bibfield  {journal} {\bibinfo
  {journal} {J. Mater. Chem.}\ }\textbf {\bibinfo {volume} {21}},\ \bibinfo
  {pages} {15843} (\bibinfo {year} {2011})}\BibitemShut {NoStop}%
\bibitem [{\citenamefont {Nolas}\ \emph {et~al.}(1998)\citenamefont {Nolas},
  \citenamefont {Cohn}, \citenamefont {Slack},\ and\ \citenamefont
  {Schujman}}]{nolas_semiconducting_1998}%
  \BibitemOpen
  \bibfield  {author} {\bibinfo {author} {\bibfnamefont {G.~S.}\ \bibnamefont
  {Nolas}}, \bibinfo {author} {\bibfnamefont {J.~L.}\ \bibnamefont {Cohn}},
  \bibinfo {author} {\bibfnamefont {G.~A.}\ \bibnamefont {Slack}}, \ and\
  \bibinfo {author} {\bibfnamefont {S.~B.}\ \bibnamefont {Schujman}},\ }\href
  {\doibase 10.1063/1.121747} {\bibfield  {journal} {\bibinfo  {journal} {Appl.
  Phys. Lett.}\ }\textbf {\bibinfo {volume} {73}},\ \bibinfo {pages} {178}
  (\bibinfo {year} {1998})}\BibitemShut {NoStop}%
\bibitem [{\citenamefont {Cohn}\ \emph {et~al.}(1999)\citenamefont {Cohn},
  \citenamefont {Nolas}, \citenamefont {Fessatidis}, \citenamefont {Metcalf},\
  and\ \citenamefont {Slack}}]{cohn_glasslike_1999}%
  \BibitemOpen
  \bibfield  {author} {\bibinfo {author} {\bibfnamefont {J.~L.}\ \bibnamefont
  {Cohn}}, \bibinfo {author} {\bibfnamefont {G.~S.}\ \bibnamefont {Nolas}},
  \bibinfo {author} {\bibfnamefont {V.}~\bibnamefont {Fessatidis}}, \bibinfo
  {author} {\bibfnamefont {T.~H.}\ \bibnamefont {Metcalf}}, \ and\ \bibinfo
  {author} {\bibfnamefont {G.~A.}\ \bibnamefont {Slack}},\ }\href {\doibase
  10.1103/PhysRevLett.82.779} {\bibfield  {journal} {\bibinfo  {journal} {Phys.
  Rev. Lett.}\ }\textbf {\bibinfo {volume} {82}},\ \bibinfo {pages} {779}
  (\bibinfo {year} {1999})}\BibitemShut {NoStop}%
\bibitem [{\citenamefont {Blake}\ \emph {et~al.}(1999)\citenamefont {Blake},
  \citenamefont {M{\o}llnitz}, \citenamefont {Kresse},\ and\ \citenamefont
  {Metiu}}]{blake_why_1999}%
  \BibitemOpen
  \bibfield  {author} {\bibinfo {author} {\bibfnamefont {N.~P.}\ \bibnamefont
  {Blake}}, \bibinfo {author} {\bibfnamefont {L.}~\bibnamefont {M{\o}llnitz}},
  \bibinfo {author} {\bibfnamefont {G.}~\bibnamefont {Kresse}}, \ and\ \bibinfo
  {author} {\bibfnamefont {H.}~\bibnamefont {Metiu}},\ }\href {\doibase
  10.1063/1.479615} {\bibfield  {journal} {\bibinfo  {journal} {J. Chem.
  Phys.}\ }\textbf {\bibinfo {volume} {111}},\ \bibinfo {pages} {3133}
  (\bibinfo {year} {1999})}\BibitemShut {NoStop}%
\bibitem [{\citenamefont {Dong}\ \emph {et~al.}(2000)\citenamefont {Dong},
  \citenamefont {Sankey}, \citenamefont {Ramachandran},\ and\ \citenamefont
  {McMillan}}]{dong_chemical_2000}%
  \BibitemOpen
  \bibfield  {author} {\bibinfo {author} {\bibfnamefont {J.}~\bibnamefont
  {Dong}}, \bibinfo {author} {\bibfnamefont {O.~F.}\ \bibnamefont {Sankey}},
  \bibinfo {author} {\bibfnamefont {G.~K.}\ \bibnamefont {Ramachandran}}, \
  and\ \bibinfo {author} {\bibfnamefont {P.~F.}\ \bibnamefont {McMillan}},\
  }\href {\doibase 10.1063/1.373447} {\bibfield  {journal} {\bibinfo  {journal}
  {J. Appl. Phys.}\ }\textbf {\bibinfo {volume} {87}},\ \bibinfo {pages} {7726}
  (\bibinfo {year} {2000})}\BibitemShut {NoStop}%
\bibitem [{\citenamefont {Dong}, \citenamefont {Sankey},\ and\ \citenamefont
  {Myles}(2001)}]{dong_theoretical_2001}%
  \BibitemOpen
  \bibfield  {author} {\bibinfo {author} {\bibfnamefont {J.}~\bibnamefont
  {Dong}}, \bibinfo {author} {\bibfnamefont {O.~F.}\ \bibnamefont {Sankey}}, \
  and\ \bibinfo {author} {\bibfnamefont {C.~W.}\ \bibnamefont {Myles}},\ }\href
  {\doibase 10.1103/PhysRevLett.86.2361} {\bibfield  {journal} {\bibinfo
  {journal} {Phys. Rev. Lett.}\ }\textbf {\bibinfo {volume} {86}},\ \bibinfo
  {pages} {2361} (\bibinfo {year} {2001})}\BibitemShut {NoStop}%
\bibitem [{\citenamefont {Tse}, \citenamefont {Li},\ and\ \citenamefont
  {Uehara}(2001)}]{tse_phonon_2001}%
  \BibitemOpen
  \bibfield  {author} {\bibinfo {author} {\bibfnamefont {J.~S.}\ \bibnamefont
  {Tse}}, \bibinfo {author} {\bibfnamefont {Z.}~\bibnamefont {Li}}, \ and\
  \bibinfo {author} {\bibfnamefont {K.}~\bibnamefont {Uehara}},\ }\href
  {\doibase 10.1209/epl/i2001-00515-8} {\bibfield  {journal} {\bibinfo
  {journal} {Europhys. Lett.}\ }\textbf {\bibinfo {volume} {56}},\ \bibinfo
  {pages} {261} (\bibinfo {year} {2001})}\BibitemShut {NoStop}%
\bibitem [{\citenamefont {He}\ and\ \citenamefont
  {Galli}(2014)}]{he_nanostructured_2014}%
  \BibitemOpen
  \bibfield  {author} {\bibinfo {author} {\bibfnamefont {Y.}~\bibnamefont
  {He}}\ and\ \bibinfo {author} {\bibfnamefont {G.}~\bibnamefont {Galli}},\
  }\href {\doibase 10.1021/nl501021m} {\bibfield  {journal} {\bibinfo
  {journal} {Nano Lett.}\ }\textbf {\bibinfo {volume} {14}},\ \bibinfo {pages}
  {2920} (\bibinfo {year} {2014})}\BibitemShut {NoStop}%
\bibitem [{\citenamefont {Mingo}\ \emph {et~al.}()\citenamefont {Mingo},
  \citenamefont {Stewart}, \citenamefont {Broido}, \citenamefont {Lindsay},\
  and\ \citenamefont {Li}}]{mingo_ab_2014}%
  \BibitemOpen
  \bibfield  {author} {\bibinfo {author} {\bibfnamefont {N.}~\bibnamefont
  {Mingo}}, \bibinfo {author} {\bibfnamefont {D.~A.}\ \bibnamefont {Stewart}},
  \bibinfo {author} {\bibfnamefont {D.~A.}\ \bibnamefont {Broido}}, \bibinfo
  {author} {\bibfnamefont {L.}~\bibnamefont {Lindsay}}, \ and\ \bibinfo
  {author} {\bibfnamefont {W.}~\bibnamefont {Li}},\ }in\ \href
  {http://link.springer.com/chapter/10.1007/978-1-4614-8651-0_5} {\emph
  {\bibinfo {booktitle} {Length-Scale Dependent Phonon Interactions}}},\
  \bibinfo {series and number} {\bibinfo {series} {Topics in Applied Physics}\
  No.\ \bibinfo {number} {128}},\ \bibinfo {editor} {edited by\ \bibinfo
  {editor} {\bibfnamefont {S.~L.}\ \bibnamefont {Shindé}}\ and\ \bibinfo
  {editor} {\bibfnamefont {G.~P.}\ \bibnamefont {Srivastava}}}\ (\bibinfo
  {publisher} {Springer New York})\ pp.\ \bibinfo {pages} {137--173},\ \bibinfo
  {note} {{DOI}: 10.1007/978-1-4614-8651-0\_5}\BibitemShut {NoStop}%
\end{thebibliography}
%

\end{document}